# Built-in Electric-Field-Control of Magnetic Coupling in van der Waals semiconductors

Chengxi Huang[1,2], Jingtong Guan[1,2], Qiongyu Li[1], Fang Wu[3], Puru Jena[2*], Erjun Kan[1*]

[1] *Department of Applied Physics and Institution of Energy and Microstructure, Nanjing University of Science and Technology, Nanjing, Jiangsu 210094, P. R. China*

[2] *Department of Physics, Virginia Commonwealth University, Richmond, Virginia 23284, United States*

[3] *College of Information Science and Technology, Nanjing Forestry University, Nanjing, Jiangsu 210037, P. R. China.*

\* Correspondence and requests for materials should be addressed to

E. K. (ekan@njust.edu.cn), P. J. (pjena@vcu.edu)

**Abstract**

Electrical control of magnetism in a two-dimensional (2D) semiconductor is of great interest for emerging nanoscale low-dissipation spintronic devices. Here, we propose a general approach of tuning magnetic coupling and anisotropy of a van der Waals (vdW) 2D magnetic semiconductor via a built-in electric field generated by the adsorption of superatomic ions. Using first-principles calculations, we predict a significant enhancement of ferromagnetic (FM) coupling and a great change of magnetic anisotropy in 2D semiconductors when they are sandwiched between superatomic cations and anions. The magnetic coupling is directly affected by the built-in electric field, which lifts the energy levels of mediated ligands' orbitals and enhances the super-exchange interactions. These findings will be of interest for ionic gating controlled ferromagnets and magnetoelectronics based on vdW 2D semiconductors.

Electric-field control of magnetism is a fundamental physical issue which has the potential to achieve the long-sought goal of high-efficient and low-dissipation spintronic devices, such as non-volatile magnetic memories [1-3]. Previous studies have proposed various strategies of manipulating magnetism of bulk and thin film systems such as diluted magnetic semiconductors [4,5], ferroelectric-ferromagnet interfaces [6,7], ferromagnetic (FM) metal films [8,9] and multiferroic materials [10-13]. These studies can be roughly divided into three areass: i) Modulating the magnitude of magnetization, which is mainly based on magnetic metals. ii) Rotating the direction of magnetization via tuning the magnetic anisotropy. iii) Transforming the magnetic ordering, e.g. from antiferromagnetic (AFM) to FM order, through tailoring the magnetic coupling between spins.

The recent discovery of van der Waals (vdW) two-dimensional (2D) magnetic semiconductors [14,15] has provided a brand-new playground for exploring unusual electrical control of magnetism. For instance, the interlayer magnetic coupling of vdW bilayers was tunable by a vertical gate voltage [16-18]. Magnetization can be rotated by a 120° ferroelectric switching in a multiferroic monolayer [19]. Magnetic phase transition was realized by asymmetric ferroelectric switching [20]. Magnetic anisotropy can also be tuned by switching the electric polarization of a ferroelectric-ferromagnetic vdW heterostructure [21]. However, a direct coupling between magnetic order and electric field has yet to be revealed in a vdW monolayer. Besides, because of the instability of 2D magnetic order, the magnetoelectric effect can only be observed at very low temperature.

In this letter, we propose a general mechanism of realizing direct electrical control of magnetic coupling, which can be applied to all kinds of normal vdW 2D magnetic semiconductors. Using a tight-binding model based on a $CrI_3$-like system, we first demonstrate that a strong vertical electric field will lift the energy level of *p* orbitals and enhance the *p-d* interaction, which results in a significant enhancement of FM super-exchange. Then, through first-principles calculations, we predict that such vertical electric field can be built in by adsorbing equal amount of superatomic cations and anions on opposite sides of a 2D semiconductor, which is controllable by an ionic gate. By this way, the FM couplings of a series of vdW 2D magnetic semiconductors are enhanced with Curie temperautre ($T_C$) increased by 25~120%, up to room temperature (300 K). Furthermore, an AFM-to-FM phase transition and a great change of magnetic anisotropy is also realized.

Our first-principles calculations are based on density functional theory (DFT) implemented in the Vienna Ab initio Simulation Package [22]. Generalized gradient approximation for exchange-correlation functional given by Perdew, Burke, and Ernzerhof [23] is used. The effective Hubbard $U_{eff}$ = 3 is added according to Dudarev's [24] method for the Cr-*d* and Mn-*d* orbitals, respectively. The projector augmented wave [25] method is used to treat the core electrons. The cutoff energy for the plane wave is set to 500 eV. The first Brillouin zone is sampled by using a Γ-centered 8 × 8 × 1 Monkhorst-Pack [26] grid. For the calculations of magnetic anisotropic energy (MAE = $E_{\text{out-of-plane}}$ − $E_{\text{in-plane}}$) per metal atom which include spin-orbit coupling (SOC) effects, a K-mesh of 14 × 14 × 1 is used. A vacuum space of 30 Å along the c axis is adopted to model the 2D system. The convergence criteria for energy and Hellmann-Feynman force components are set to 1×10$^{−5}$ eV and 0.01 eV/Å, respectively. The van der Waals correction is incorporated using the DFT-D2 method [27]. The dipole-dipole interaction corrections [28] are also included.

To explore how a vertical electric field affects the magnetic coupling of a monolayer system, here we introduce a tight-binding double-orbital model to mimic the magnetic interactions in a typical 2D FM semiconductor, e.g. CrI$_3$ monolayer (Fig. 1a). This model was sucessuful in explaining the strong magnetic couplings of alloyed transition metal compounds [29]. To take into account the effect of *p* orbitals of two I atomic layers, we include two degenerate non-mangetic orbitals. Based on the mean field treatment, the general Hamiltonian of the magnetic cluster $\alpha$-$\gamma_{I,II}$-$\beta$ is written as [30]

$$\hat{H} = \sum_{li} \epsilon_{li} \hat{d}^{\dagger}_{li} \hat{d}_{li} + \sum_{\gamma k} \epsilon_{\gamma k} \hat{p}^{\dagger}_{\gamma k} p_{\gamma k} + \sum_{li,\gamma k} [t_{li,\gamma k} \hat{d}^{\dagger}_{li} p_{\gamma k} + h.c.]$$

$$+ \sum_{\alpha i,\beta j} [t_{\alpha i,\beta j} \hat{d}^{\dagger}_{\alpha i} \hat{d}_{\beta j} + h.c.] + \frac{U}{2} \sum_{l} \vec{e}_l \cdot \vec{S}_l$$

where $l = \alpha, \beta$ represent spin-polarization orbitals (e.g. $d$ orbitals) for magnetic ions and $\gamma_{I,II}$ represent orbitals (e.g. $p$ orbitals) of non-magnetic ligand ions. The first and second terms denote the on-site energies of each orbital. The third term is the hopping term between magnetic and non-magnetic orbitals. The 4th term is the direct hopping between neighboring magnetic ions. The last term denotes the exchange field for magnetic ions. The interaction between the two non-magnetic ligands' orbitals is omitted here because it barely affects the results. A crystal field splitting ($\Delta_c$) of the magnetic ions is introduced to mimic a semiconducting system (Fig. 1b).

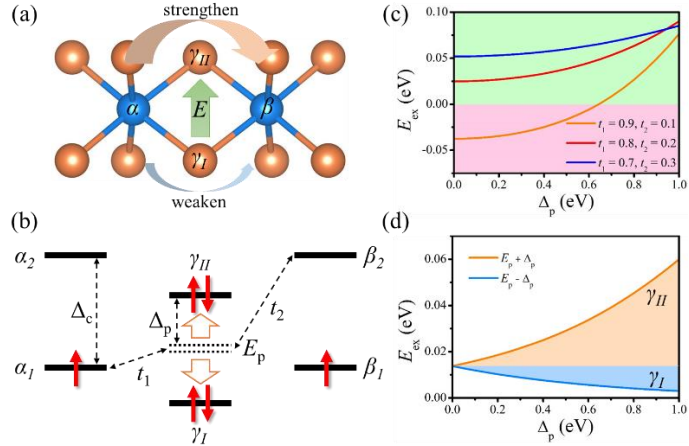

**Fig. 1**. (a) A magnetic cluster including two nearest-neighbor magnetic sites in a CrI$_3$-like material. Orange and blue balls represent non-magnetic ligand ions (e.g. I) and magnetic ions (e.g. Cr), respectively. (b) Schematic diagram of the double-orbital model. $\alpha$ and $\beta$ orbitals are spin-polarized orbitals of magnetic ions. $\gamma$ orbitals are fully occupied orbitals of non-magnetic ions. (c) The exchange energy ($E_{ex} = E_{AFM} - E_{FM}$) as a function of the on-site energy level split of the two non-magnetic orbitals ($\Delta_p$) when the hopping strength $t_1 > t_2$. (d) The individual influences of the raising of $\gamma_{II}$ orbital and the lowering of $\gamma_I$ orbital on the $E_{ex}$. $t_1$ and $t_2$ are set to 0.7 and 0.3 eV, respectively. The crystal field splitting ($\Delta_c$), exchange field ($U$), initial on-site energy of $\gamma$ orbitals ($E_p$) and direct hopping strength between $\alpha$ and $\beta$ orbitals ($t_d$) are set to 2, 4, 0.2 (the on-site energies of $\alpha_1$ and $\beta_1$ orbitals are set to zero) and 0.05 eV, respectively.

It can be easily demonstrated by this model that increasing the hopping strength between $\alpha$ ($\beta$) and $\gamma$ orbitals (i.e. $t_1$ and $t_2$) or raising the on-site energy levels of $\gamma$ orbitals will enhance the FM coupling strength (Fig. S1 [31]). This usually depends on the choice of the ligands. For instance, changing the ligands (X) of a CrX$_3$ (X = Cl, Br, I) monolayer from Cl to I will distinctly increase its $T_C$ [34], because the on-site energy level of I-$p$ orbitals are relatively higher than that of Cl-$p$ orbitals in the CrX$_3$ systems. This can be understood as follows: increasing the covalency, namely the $d$-$p$ hybridization, will enhance the FM super-exchange interactions in a CrX$_3$-like system.

Applying a vertical electric field will induce a potential difference between the two ligand atomic layers, namely, an energy level split ($\Delta_p$) of the two degenerate $\gamma$ orbitals. The energy level of the $\gamma_{II}$ orbital is raised but that of the $\gamma_I$ orbital is lowered. Thus, to explore how $\Delta_p$ affects the magnetic coupling, we calculated the exchange energies ($E_{ex} = E_{AFM} - E_{FM}$, where $E_{AFM}$ and $E_{FM}$ represent the energy of AFM and FM states, respectively.) of the model system with different $\Delta_p$ via direct diagonalization of the Hamiltonian matrix. Interestingly, the results (Fig. 1c) show that the $E_{ex}$ monotonically increases with increasing $\Delta_p$ and larger the ratio $t_1/t_2$ the steeper is the profile (see Section 1 [31] for more details). To better understand the underlying mechanism of the enhancement of FM coupling, we separately considered the effects of $\gamma_I$ and $\gamma_{II}$ orbitals (Fig. 1d and Fig. S3 [31]). As expected, the raising of $\gamma_{II}$ orbital increases the $E_{ex}$, whereas the lowering of $\gamma_I$ orbital decreases it. However, the $\gamma_{II}$ orbital wins the competition and dominates the influence on magnetic coupling when $\Delta_p$ is large. Thus, a vertical electric field, which induces a large $\Delta_p$, will distinctly enhance the FM coupling of a CrI$_3$-like system. It is worth noting that, the increase of $\Delta_p$ may also result in an AFM-to-FM phase transition (Fig. 1c).

We first tried directly applying a vertical electric field on the CrI$_3$ monolayer. However, the DFT results show that when the electric field is 5 V/nm, the magnetic coupling is barely affected ($E_{ex}$ changes from 26 to 27 meV per Cr). This is because the $\Delta_p$ is too small (only ~0.1 eV from the planar average of electrostatic potential energy shown in left panel of Fig. 2c) due to the electronic screening.

An alternative way to introduce a large potential energy difference is through edge or surface modification using different functional groups [35]. But surface modification using traditional functional groups may cause surface reconstruction and destroy the intrinsic property of 2D materials. Therefore, here we choose superatomic ions [36], which mimic the chemistry of elemental ions but could usually remain intact when adsorbed on or embedded into materials. To avoid additional valence electrons, we use two different superatomic ions with opposite valence states, namely the B(CN)$_4^-$ and NH$_4^+$, placed on opposite sides of the CrI$_3$ surface (see Fig. 2a). We denote this sandwich-like structure as B(CN)$_4$/Cr$_2$I$_6$/NH$_4$. In this case, the superatomic anions and cations behave as electron acceptors and donors, while the number of valence electrons in the CrI$_3$ layer does not change. One electron may transfer from NH$_4$ to B(CN)$_4$ across the CrI$_3$ layer, forming NH$_4^+$ and B(CN)$_4^-$ and generating a built-in electric field, which may induce a large $\Delta_p$.

Different possible adsorption sites were considered to determine the optimal geometry of B(CN)$_4$/Cr$_2$I$_6$/NH$_4$ (see Section 2 [31]). The results show that both B(CN)$_4^-$ and NH$_4^+$ preferably adsorb on the Cr-top sites. After optimization, the CrI$_3$ framework is well maintained except for a small shift of Cr ions along the *c* direction. The structural symmetry is reduced from $D_{3d}$ (for pristine CrI$_3$ monolayer) to $C_3$. The adsorption energy (defined as $E_a = E_{B(CN)_4/Cr_2I_6/NH_4} - E_{B(CN)_4} - E_{NH_4} - E_{Cr_2I_6}$, where $E_{B(CN)_4}$ and $E_{NH_4}$ denote the total energies of isolated $B(CN)_4$ and $NH_4$ clusters, respectively) is calculated to be -4.79 eV. The adsorption distances (defined as the shortest perpendicular distance between the superatomic ion and the 2D layer) for $B(CN)_4$ and $NH_4$ on $CrI_3$ layer are 2.35 and 1.94 Å, respectively, which are much smaller than the common vdW adsorption distance for molecules (e.g. ~3.03 and 3.22 Å for $CO_2$ and $N_2$ molecules, respectively, when adsorbed on $CrGeTe_3$ surface) [37]. These imply the formation of strong ionic bonds. To confirm the thermal stability of this sandwich structure, *ab* initio molecular dynamics simulations were also performed at 300 K (see Section 3 [31]).

A possible approach to realize such a sandwich structure is to use the ionic liquid gating (Fig. 2b), which can be controlled by a gate voltage and is a widely adopted approach to manipulate magnetism of materials [38-43]. In particular, it has been possible to raise the $T_C$ of metallic $Fe_3GeTe_2$ vdW thin flakes to room temperature via electrostatic carrier doping through an ionic gate [38]. In our case, a gate voltage across the ionic liquid will separate the cations and anions and result in a selective adsorption on the $CrI_3$ surface. Besides, the applied gate voltage can control the concentration of the adsorbed superatomic ions.

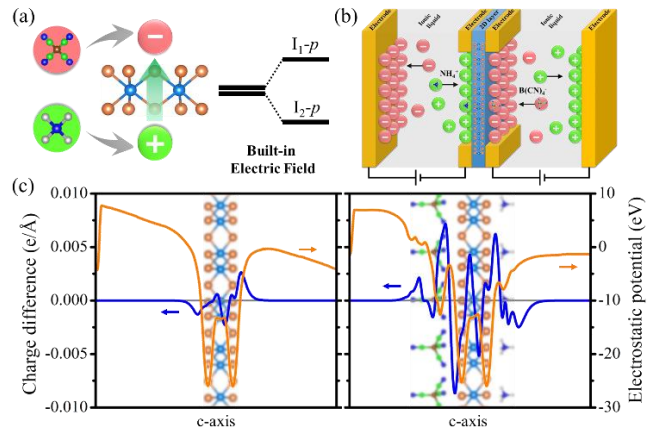

**Fig. 2**. (a) Schematic diagram of inducing a splitting of *p* orbitals through a built-in electric field formed by the adsorption of superatomic ions. (b) A possible approach

to realize the sandwich structure built by a vdW 2D semiconductor and superatomic ions via the ionic liquid gating. The vdW 2D FM semiconductor (e.g. $CrI_3$) is placed in the middle. By applying a positive gate voltage across the ionic liquid containing $NH_4^+$ (e.g. $NH_4Cl$), the $NH_4^+$ will adsorb on the left side of $CrI_3$ layer. On the contrary, a negative gate voltage applied across the ionic liquid (e.g. [EMIM][B(CN)$_4$]) will make the $B(CN)_4^-$ adsorb on the right side of $CrI_3$ layer. (c) Planar average charge difference and electrostatic potential of $CrI_3$ monolayer under an external out-of-plane electric field (left panel) and under a built-in electric field generated by the adsorbed superatomic ions [$B(CN)_4^-$ and $NH_4^+$] (right panel). Insets: the side view of $CrI_3$ monolayer and $B(CN)_4/Cr_2I_6/NH_4$. Light blue, orange, brown, green, dark blue, gray and cyan balls represent Cr, I/Te, N, C, N, H and Ge atoms, respectively.

Next, we explore the electronic and magnetic properties of $B(CN)_4/Cr_2I_6/NH_4$. Same as in the pristine $CrI_3$ monolayer, the spin-polarization is mainly contributed by the Cr-$d$ orbitals (Fig. S10 [31]) with a formal magnetic moment of 3 $\mu_B$ per Cr. The adsorption of superatomic ions brings three nearly non-spin-polarized flat bands intercalated into the energy gap of $CrI_3$ (Fig. 3a and b), which are mainly contributed by the $B(CN)_4^-$ (Fig. S11a [31]). This reduces the electronic band gap from 1.08 to 0.22 eV (Fig. 3b). As expected, a distinct charge rearrangement of the $CrI_3$ layer occurs due to the adsorption of $B(CN)_4^-$ and $NH_4^+$ (right panel in Fig. 2c and Fig. S11b). The Bader charge analysis [44] shows a large amount of charge (~0.7) transferring from $NH_4$ to $B(CN)_4$ across the $CrI_3$ layer. There is also a small charge difference (~0.3) for the two I atomic layers. Interestingly, a distinct split of $p$ orbitals for the two I atomic layers is observed (Fig. S11c [31]) with a $\Delta_p$ of ~0.67 eV (right panel in Fig. 2c). This implies an extremely strong built-in electric field. As a result, the ground state of $B(CN)_4/Cr_2I_6/NH_4$ is FM with $E_{ex}$ of 60 meV per Cr, more than double compared to the value of 26 meV for the pristine $CrI_3$ monolayer.

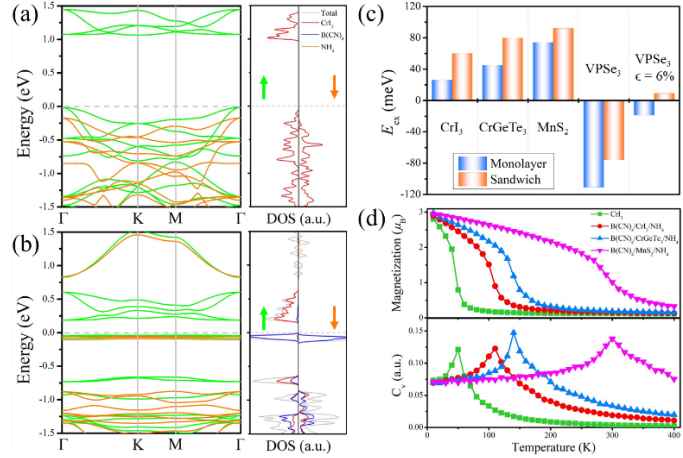

**Fig. 3**. Spin-resolved band structure and projected density of states (DOS) of (a) pristine $CrI_3$ monolayer and (b) $B(CN)_4/Cr_2I_6/NH_4$. Green and orange profiles represent spin-up and spin-down bands, respectively. (c) The exchange energies ($E_{ex}$ = $E_{AFM}$ - $E_{FM}$) for different magnetic semiconductor monolayers before (blue cylinders) and after (orange cylinders) adsorbing $B(CN)_4^-$ and $NH_4^+$ on their two sides. Positive and negative values represent ferromagnetic and antiferromagnetic couplings, respectively. (d) Magnetic moment per metal atom and specific heat ($C_v$) as function of temperature for pristine $CrI_3$ monolayer, $B(CN)_4/Cr_2I_6/NH_4$, $B(CN)_4/Cr_2Ge_2Te_6/NH_4$ and $B(CN)_4/Mn_4S_8/NH_4$.

To further understand the mechanism of the remarkable enhancement of FM coupling for $B(CN)_4/Cr_2I_6/NH_4$, we have also considered the effect of structural distortion (including the change of $d$ energy levels and $d$-$d$ direct-exchange interactions) and the extra super-exchange paths provided by the adsorbed molecules, which are proved not to be the main factors (see Section 4 [31] for details). In addition, when we reduce the adsorption concentration of the superatomic ions, the $E_{ex}$ quickly decreases from 60 to 22 meV per Cr (see Section 2 [31] for details). This can be understood by knowing that the strength of the built-in electric field is proportional to the adsorption concentration of superatomic ions. A lower adsorption concentration leads to a smaller built-in electric field and a smaller $E_{ex}$. Similarly, if we apply an external electric field of 5 and 10 V/nm on B(CN)$_4$/Cr$_2$I$_6$/NH$_4$, antiparallel to the built-in electric field, the $E_{ex}$ will decrease from 60 to 53 and 45 meV per Cr, respectively. These results are consistent with our model analysis (Fig. 1c). Thus, through above discussions we can conclude that the enhancement of FM coupling in B(CN)$_4$/Cr$_2$I$_6$/NH$_4$ is indeed dominated by the strong built-in electric field induced by the adsorption of superatomic ions.

This mechanism can also be applied to other similar vdW magnetic monolayers. To demonstrate this, we studied a series of sandwich systems built by different vdW 2D magnetic semiconductors (i.e. CrI$_3$, CrGeTe$_3$, MnS$_2$ and VPSe$_3$) and superatomic ions [i.e. B(CN)$_4^-$, BF$_4^-$, NH$_4^+$, Li$_3$O$^+$ and Na$_3$O$^+$] (see Fig. S13 and table S1 [31]). The results show that, different combination of adsorbed superatomic cations and anions results in different electronic band gaps, which mainly depends on the location of the localized energy bands of the superatomic anions around the Fermi level (Fig. 3a, b and Fig. S14 [31]). For CrI$_3$, CrGeTe$_3$ and MnS$_2$, the adsorption of superatomic ions will distinctly enhance their FM coupling. While for VPSe$_3$, the adsorption of superatomic ions will weaken its AFM coupling. Interestingly, for VPSe$_3$ under an in-plane tensile strain of 6%, the adsorption of superatomic ions leads to an AFM-to-FM phase transition.

The adsorption of superatomic ions not only affects the magnetic coupling, but also greatly changes the magneto-crystalline anisotropy. For CrI$_3$, the MAE changes from -0.57 to 1.74 meV (negative and positve values represnet out-of-plane and in-plane magnetic easy axis, respectively) after adsorbing B(CN)$_4^-$ and NH$_4^+$. On the other hand, for CrGeTe$_3$-based systems, the MAE changes from -0.3 to -1.25 meV. Previous works have pointed out that, according to the perturbation theory, magnetic anisotropy for 2D CrI$_3$ is mainly contributed by the I-$p$ orbitals near the Fermi level through indirect SOC between Cr and I atoms [45-47]. In addition, the atomic SOC effects from the light elements (e.g. B, C, N and H) in the superatomic ions are very weak and can be neglected. Thus the sign of MAE, namely, the direction of easy axis is determined by $\sigma\sigma'(|<o, \sigma|L_z|u, \sigma'>|^2 - |<o, \sigma|L_x|u, \sigma'>|^2)$ (where $o$ and $u$ represent occupied and unoccupied states of I atoms, respectively, $\sigma$ and $\sigma'$ are spin indices). From the projected density of states for CrI$_3$ (Fig. S15a [31]), we see that the $o$ and $u$ states near the Fermi level are dominated by $p$x and $p$y states in the same spin channel ($\sigma\sigma' = 1$), leading to an out-of-plane easy axis. While for the sandwiched B(CN)$_4$/Cr$_2$I$_6$/NH$_4$, the occupied I-$p$z states rise near the Fermi level (Fig. S15b [31]), resulting in an increase of $|<o, \sigma|L_x|u, \sigma'>|$ term between $p$x/$p$y and $p$z (different magnetic quantum number), which causes a sign reversal of MAE. For CrGeTe$_3$ and B(CN)$_4$/Cr$_2$Ge$_2$Te$_6$/NH$_4$, the electronic states around the Fermi level are both dominated by Te-$p$x,$p$y states (Fig. R5c and d). Thus, the sign of MAE does not change. The increase of absolute value of MAE caused by the adsorption of superatomic ions (from 0.57 to 1.74 meV for CrI$_3$ and from 0.3 to 1.25 meV for CrGeTe$_3$) can be understood that the enhancement of $d$-$p$ interaction induced by the built-in electric field makes the indirect SOC more efficient.

To directly show how the adsorption of superatomic ions affects the streght of FM coupling for the studied systems, we estimate their $T_C$ by performing Metropolis Monte Carlo (MC) simulations [48] based on the Heisenberg model including single-ion anisotropy (see section 5 [31] for details). The spin Hamiltonian is written as

$$\hat{H} = -\sum_{<ij>} J\vec{S_i} \cdot \vec{S_j} + \sum_{<i>} A(S_{iz})^2,$$

where $J$ is the nearest neighbor exchange interaction parameter and the summation <ij> runs over all the nearest neighbor Cr sites. $A$ is the single-ion anisotropic parameter, $S_{iz}$ represent components of $S$ along $z$ (out-of-plane) orientation. The evolution of magnetization and specific heat of pristine CrI$_3$ monolayer, B(CN)$_4$/Cr$_2$I$_6$/NH$_4$, B(CN)$_4$/Cr$_2$Ge$_2$Te$_6$/NH$_4$, and B(CN)$_4$/Mn$_4$S$_8$/NH$_4$ along with the increase in temperature is shown in Fig. 3d. The estimated $T_C$ for all the considered sandwich systems are listed in Table S1 [31], which are increased by 25~120% compared to that of the pristine monolayers. Among them, the $T_C$ of $B(CN)_4/Mn_4S_8/NH_4$ and $B(CN)_4/Mn_4S_8/Li_3O$ reach room temperature (300 K). Note that the $B(CN)_4/Cr_2I_6/NH_4$ shows an in-plane easy axis anisotropy. Thus a small external magnetic field might be needed to stabilize its long-range FM order below $T_C$ (see section 5 [31] for details).

In summary, using a general tight-binding model and first-principles calculations, we have revealed a direct coupling between magnetic order and electric field in vdW 2D magnetic semiconductors. The vertical electric field lifts the energy levels of mediated ligands' orbitals and enhances the super-exchange interaction, leading to a significant increase of $T_C$ and an AFM-to-FM phase transition. This mechanism is made possible by sandwiching the 2D magnetic semiconductor between superatomic cations and anions through ionic gating, which induces a strong built-in electric field. We arrived at this conclusion by considering four 2D magnetic systems, namely, $CrI_3$, $CrGeTe_3$, $MnS_2$ and $VPSe_3$ monolayers, sandwiched by different superatomic cations and anions. The $T_C$ increased by 25~120% upon forming the sandwich complexes and up to room temperature (300 K) in $MnS_2$-based systems. For $VPSe_3$-based system, an AFM-to-FM phase transition was realized. Besides, the adsorption of superatomic ions can also tune the magnetic anisotropy. We look forward to experimental realization of manipulating magnetism of vdW 2D magnetic semiconductors via adsorption of superatomic ions controlled by ionic liquid gating.

The work is supported by the NSFC (51522206, 11774173, 11474165), by the Youth Program of NSFC (12004183), by the Fundamental Research Funds for the Central Universities (No.30915011203), and by the Outstanding Youth Fund of Nanjing Forestry University (NLJQ2015-03). C.H. and E.K. acknowledge the support from the Tianjing Supercomputer Centre and Shanghai Supercomputer Center. P.J. acknowledges partial supported from the U.S. Department of Energy, Office of Basic Energy Sciences, Division of Materials Sciences and Engineering under Award No.DE-FG02-96ER45579. We thank Hongjun Xiang from Fudan University, Jian Zhou from Xi'an Jiaotong University and Hong Fang from Virginia Commonwealth University for valuable discussions.

corresponding electronic properties, stabilities, details of Monte Carlo simulations and properties for different magnetic sandwich systems.